\documentclass[aps,showpacs,nofootinbib,superscriptaddress]{revtex4-1}
\usepackage{graphicx}
\usepackage{dcolumn}

\begin{document}

\title{ Study of the strong  $\Sigma_b\to \Lambda_b\, \pi$ and 
$\Sigma_b^{*}\to \Lambda_b\, \pi$  in a non-relativistic quark model}

\author{ E. Hern\'andez} \affiliation{Departamento de F\'\i sica Fundamental e
IUFFyM, Universidad de Salamanca, E-37008 Salamanca, Spain.}  
\author{J.~Nieves} \affiliation{Instituto de F\'\i sica Corpuscular
(IFIC), Centro Mixto CSIC-Universidad de Valencia, Institutos de
Investigaci\'on de Paterna, Aptd. 22085, E-46071 Valencia, Spain}
\begin{abstract} 
\rule{0ex}{3ex} 

\end{abstract}


\begin{abstract} 
  We present results for the strong widths corresponding to the
 $\Sigma_b\to \Lambda_b\, \pi$ and $\Sigma_b^{*}\to \Lambda_b\, \pi$
 decays. We apply our model in Ref. Phys. Rev. D 72, 094022 (2005)
 where we previously studied the corresponding transitions in the
 charmed sector. Our non-relativistic constituent quark model uses
 wave functions that take advantage of the constraints imposed by
 heavy quark symmetry. Partial conservation of axial current
 hypothesis allows us to determine the strong vertices from an
 analysis of the axial current matrix elements.
\end{abstract}
\maketitle

In this work we extend our model in Ref.~\cite{Albertus:2005zy} to the bottom 
sector by
evaluating strong one-pion decays of the $\Sigma_b$ and $\Sigma_b^*$ baryons
into  $\Lambda_b^0\,\pi$. The $\Lambda_b^0$ was first observed in 1991 by the
UA1 Collaboration at the CERN proton-antiproton collider in the
$\Lambda_b^0\to J/\Psi\Lambda$ decay channel~\cite{Albajar:1991sq}.  A precise
determination of its mass 
 was performed by the CDF Collaboration 
 in 2006 analyzing exclusive $J/\Psi$ decays~\cite{{Acosta:2005mq}}. 
 In 2007 the CDF Collaboration observed for the first time the
 $\Sigma_b^{\pm}$ and $\Sigma_b^{*\,\pm}$ baryons and their masses were also
  measured ~\cite{:2007rw}. New results on the $\Sigma_b^{(*)\,\pm}$ masses by
  the CDF Collaboration are compiled in their Public Note  10286 (See the CDF
  Collaboration web page). 
 The $\Xi_b^-$ baryon was first observed by the D0
 Collaboration~\cite{d0-07} and its mass was later measured with
 precision by the CDF Collaboration ~\cite{cdf09}. The CDF
 Collaboration has just reported the observation of its isospin
 partner the $\Xi_b^0$~\cite{cdf11}, but there is no evidence yet for
 the $\Xi_b^{*\,-},\,\Xi_b^{*\,0}$ states. New precise results for
 bottom baryons are expected from the different LHC Collaborations at
 CERN.

 Similarly to the charm quark sector,
 $\Sigma_b^{(*)}$ decays are dominated by the
 strong decay channel $\Lambda_b^0\,\pi$. These strong decays have
 been analyzed before in Ref.~\cite{hwang07}, using an approach that
 combined chiral dynamics with the MIT bag model, in
 Ref.~\cite{guo08}, where the authors used the Bethe-Salpeter
 formalism under the covariant instantaneous approximation,  in
 Refs.~\cite{Chen:2007xf,Faessler:2010zzc, Limphirat:2010zz} within
 the non-relativistic $^3P_0$ quark model, and in 
 Refs.~\cite{azizi09,azizi11,azizi11-2} using light cone QCD sum rules (LCSR).  
 There is also an estimate
 by the CDF Collaboration~\cite{:2007rw} obtained using the expected
 initial and final baryon mass differences and the heavy quark
 effective theory (HQET) approach of Ref.~\cite{Korner:1994nh}.

In our model we use the heavy quark symmetry (HQS) constrained wave
 functions that we evaluated in Ref.~\cite{albertus04} and the pion
 emission amplitude is obtained with the use of partial conservation
 of axial current hypothesis
 (PCAC). HQS~\cite{nussinov,voloshin,politzer,iw} tell us that in the
 infinite heavy quark mass limit the dynamics of the light quark
 degrees of freedom becomes independent of the heavy quark flavor and
 spin. This allows to take the light degrees of freedom with well
 defined quantum numbers which simplifies the solution of the
 three-body problem for ground state (L=0) baryons with a heavy
 quark. In Ref.~\cite{albertus04} we solved the three-body problem
 using a variational approach and our results for the spectrum agreed
 with a previous Faddeev calculation~\cite{silvestre96} using the same
 interquark potentials. 
  
   As in the charm sector the phase space available in these reactions
is very limited, and  for that reason,  we need to use precise mass
values. When available, we shall use  physical masses taken from
Ref.~\cite{pdg10}. For the $\Sigma_b^{*0}$ strong decay we shall
give an estimate of the decay width taking, following
Ref.~\cite{hwang07},
$M_{\Sigma_b^{*\,0}}=\frac12(M_{\Sigma_b^{*\,+}}+M_{\Sigma_b^{*\,-}})
$. For the $\Sigma_b^{0}$ case, corrections to the analogous relation,
 due to the electromagnetic 
interaction between the two light quarks
in the heavy baryon,  have been evaluated
in Ref.~\cite{Hwang:2008dj} using HQET 
 and  in Ref.~\cite{Guo:2008ns} in chiral perturbation theory to leading one-loop
order. Based on the known experimental data they get
$M_{\Sigma_b^{0}}=5810.5\pm2.2\,$MeV~\cite{Hwang:2008dj} and
$M_{\Sigma_b^{0}}=5810.3\pm1.9\,$MeV~\cite{Guo:2008ns}, their central values
being 1\,MeV lower than the value one would obtain from  the less accurate relation
$M_{\Sigma_b^{\,0}
}=\frac12(M_{\Sigma_b^{\,+}}+M_{\Sigma_b^{\,-}})$. To give an
estimate of the $\Sigma_b^0$ decay width we shall use the value
$M_{\Sigma_b^{0}}=5810.5\,$MeV given in Ref.~\cite{Hwang:2008dj}.
As shown below the decay width is
proportional to the cube of the final baryon momentum and a one MeV increase in
the $\Sigma_b^0$ mass induces 
 an approximate 4\% increase in the decay width. This correction can be considered as an
estimate of the uncertainties on the widths evaluation due to the uncertainties in the 
experimental baryon masses.

As mentioned, we determine the pion emission amplitude ${\cal A}_{BB'\pi}^{(s,s')}(P_{B},P_{B'})$
 through the use of PCAC that
 allows  to relate that amplitude to the matrix element of 
the divergence of the axial current.
For a strong $B\to B'\pi^+$ decay we have that\,\footnote{Note that we give the
expression corresponding to the non--pole part of the matrix element. If
 the pion pole contribution is included then the relation is given by
$
\left\langle B',\ s'\,\vec{P}'\,\left|\,q^\mu\, J_{A\,\mu}^{d\,u}(0)
\,\right|
\,B,\ s\,\vec{P}'\,\right\rangle= -i\,f_{\pi}\frac{m_{\pi}^2}{q^2-m_{\pi}^2}\ 
{\cal A}^{(s,s')}_{BB'\pi^+}(P,P')
$.}
\begin{eqnarray}
\label{np}
\left\langle B',\ s'\,\vec{P}'\,|\,q^\mu\, J_{A\,\mu}^{d\, u}(0)\,|
\,B,\ s\,\vec{P}'\,\right\rangle_{non-pole}&=& i\,f_{\pi}\ 
{\cal A}^{(s,s')}_{BB'\pi^+}(P',P')
\end{eqnarray}
where $s,\, s'$ are the third component of the spin of the $B,\, B'$
baryons in their respective center of mass systems,
$P=(E,\,\vec{P}),\, P'=( E',\,\vec{P}') $ are their respective
four--momenta and $q=P-P'$. $J_{A\,\mu}^{d\, u}(0)$ is the axial
current for the $u\to d$ transition, and $f_{\pi}=130.41
$\,MeV~\cite{pdg10} is the pion decay constant. The baryon states are
normalized as $\left\langle B, s' \vec{P}'\,| B, s \vec{P}
\right\rangle=\delta_{s,s'}\,(2\pi)^3\,2E\,
\delta(\vec{P}-\vec{P}')$.\\
%
%
%
%

%
%
%
%
For the particular case of the $\Sigma_b^{(*)\,+}\to \Lambda_b^0$ decays we have that
\begin{eqnarray}
\label{eq:amsiglam}
&&\hspace*{-.5cm}{\cal A}^{(s,s')}_{\Sigma_b^{+}\Lambda_b^0\pi^+}(P,P')=\frac{-i}{f_\pi}\ 
\left\langle \Lambda_b^0,\ s'\,\vec{P}'\,\left|\,q^\mu\, J_{A\,\mu}^{d\, u}(0)
\,\right|
\,\Sigma_b^{+},\ s\,\vec{P}\,\right\rangle_{non-pole}
=-i\frac{g_{\Sigma_b^{+}\Lambda_b^0\pi^+}}{M_{\Sigma_b^{+}}+M_{\Lambda_b^0}}
q_\nu\ \overline{u}_{\Lambda_b^0\ s'}(\vec{P}'\,)\
\gamma^\nu\gamma_5\
u_{\Sigma_b^{+}\ s}(\vec{P})\\
\label{eq:amsig*lam}
&&\hspace*{-.5cm}{\cal A}^{(s,s')}_{\Sigma_b^{*\,+}\Lambda_b^0\pi^+}(P,P')=\frac{-i}{f_\pi}
\left\langle \Lambda_b^0,\ s'\,\vec{P}'\,\left|\,q^\mu\, J_{A\,\mu}^{d\, u}(0)
\,\right|
\,\Sigma_b^{*\,+},\ s\,\vec{P}\,\right\rangle_{non-pole}=i
\frac{g_{\Sigma_b^{*\,+}\Lambda_b^0\pi^+}}{2M_{\Lambda_b^0}}
\ q_\nu\ \overline{u}_{\Lambda_b^0\ s'}(\vec{P}'\,)\
u_{\Sigma_b^{*\,+}\ s}^\nu(\vec{P})
\end{eqnarray}
where the $g_{\Sigma_b^{(*)\,+}\Lambda_b^0\pi^+}$ are dimensionless
strong coupling constants.  $u_{\Sigma_b^{+}\ s}(\vec{P})$ and
$u_{\Lambda_b^{0}\ s'}(\vec{P}')$ are Dirac spinors and
$u_{\Sigma_b^{*\,+}\ s}^\nu(\vec{P})$ is a Rarita-Schwinger spinor all
normalized to twice the energy.

Taking $\vec{P}=\vec{0}$ and $\vec{P}'=-|\vec{q}\,|\,\vec{k}$ in the
$z$ direction, the width is given in each case by
\begin{eqnarray}
\label{eq:gsiglam}
\Gamma(\Sigma_b^{+}\to\Lambda_b^0\pi^+)&=&
\frac{|\vec{q}\,|\,g_{\Sigma_b^{+}\Lambda_b^0\pi^+}^2}{4\pi M_{\Sigma_b^{+}} }
(E_{\Lambda_b^0}-M_{\Lambda_b^0})=
\frac{|\vec{q}\,|\,g_{\Sigma_b^{+}\Lambda_b^0\pi^+}^2}{8\pi M_{\Sigma_b^{+}}^2 }\
\left[\ (
M_{\Sigma_b^{+}}-M_{\Lambda_b^0})^2-m^2_{\pi^+}\right]
\approx\frac{|\vec{q}\,|^3\,g_{\Sigma_b^{+}\Lambda_b^0\pi^+}^2}
{8\pi M_{\Lambda_b^{0}}M_{\Sigma_b^{+}} }\
\\
\label{eq:gsig*lam}
\Gamma(\Sigma_b^{*\,+}\to\Lambda_b^0\pi^+)&=&
\frac{|\vec{q}\,|^3\,g_{\Sigma_b^{*\,+}\Lambda_b^0\pi^+}^2}{12\pi M_{\Sigma_b^{+}} }
\frac{E_{\Lambda_b^0}+M_{\Lambda_b^0}}{4M_{\Lambda_b^{0}}^2}=
\frac{|\vec{q}\,|^3\,g_{\Sigma_b^{*\,+}\Lambda_b^0\pi^+}^2}
{24\pi M_{\Sigma_b^{*\,+}}^2 }\
\frac{(M_{\Sigma_b^{*\,+}}+M_{\Lambda_b^0})^2-m^2_{\pi^+}}
{4M_{\Lambda_b^{0}}^2} 
\end{eqnarray}
The final momentum is given by  $|\vec q\,|=\frac1{2M_{\Sigma_b^+}}{\lambda^{1/2}
(M_{\Sigma_b^+}^2,M_{\Lambda_b^0}^2,m_{\pi^+}^2)}$ with
$\lambda(a,b,c)=a^2+b^2+c^2-2ab-2ac-2bc$ the K\"allen function. 
The $|\vec{q}\,|^3$ behavior makes the widths  very sensitive to the
actual baryon masses used. That is the reason to use experimental
masses. The test of the model comes through the evaluation of the corresponding
coupling constants.

From the PCAC relation in Eqs.~(\ref{eq:amsiglam}-\ref{eq:amsig*lam}),
taking $\vec{P}=\vec{0}$ and $\vec{P}'=-|\vec{q}\,|\,\vec{k}$ in the
$z$ direction, and $s=s'=1/2$, we have
\begin{eqnarray}
\hspace{-.5cm}g_{\Sigma_b^{+}\Lambda_b^0\pi^+}
&=&\frac{-1}{f_\pi}\ 
\frac{\sqrt{E_{\Lambda_b^+}+M_{\Lambda_b^0}}
}{|\vec{q}\,|
\sqrt{2M_{\Sigma_b^{+}} }}
\left[
(M_{\Sigma_b^{+}}- E_{\Lambda_b^0} )\ A_{\Sigma_b^{+}\Lambda_b^0
,\ 0}^{1/2,1/2}
+|\vec{q}\,|\ A_{\Sigma_b^{+}\Lambda_b^0
,\ 3}^{1/2,1/2}
\right]\nonumber\\
\hspace{-0.5cm}g_{\Sigma_b^{*\,+}\Lambda_b^0\pi^+}
&=&\frac{\sqrt3}{f_\pi\sqrt2}\ \frac{2M_{\Lambda_b^0}
}
{|\vec{q}\,|\sqrt{2M_{\Sigma_b^{*\,+}} \left(
E_{\Lambda_b^0}+M_{\Lambda_b^0}\right)}}
\left[
(M_{\Sigma_b^{*\,+}}- E_{\Lambda_b^0} )\ A_{\Sigma_b^{*\,+}
\Lambda_b^0
,\ 0}^{1/2,1/2}
+|\vec{q}\,|\ A_{\Sigma_b^{*\,+}\Lambda_b^0
,\ 3}^{1/2,1/2}
\right]
\end{eqnarray}
with
\begin{eqnarray}
A_{\Sigma_b^{(*)\,+}\Lambda_b^0
,\ \mu}^{1/2,1/2}=\left\langle \Lambda_b^0,\
1/2\ -|\vec{q}|\,\vec{k}\ \left|\  J_{A\,\mu}^{d\, u}(0)\ \right|
\,\Sigma_b^{(*)\,+},\ 1/2\ \vec{0}\,\right\rangle_{\ non-pole}
\end{eqnarray}
The $A_{\Sigma_b^{(*)\,+}\Lambda_b^0 ,\ \mu}^{1/2,1/2}$ weak matrix
elements are easily evaluated in the model using one-body current
operators and their expressions can be found in the appendix.

In Table~\ref{tab:siglam} we present the results for
$g_{\Sigma_b^{(*)\,+}\Lambda_b^0\pi^+}$ and the widths
$\Gamma(\Sigma_b^{(*)\,+}\to\Lambda_b^0\pi^+)$,
$\Gamma(\Sigma_b^{(*)\,0}\to\Lambda_b^0\pi^0)$ and
$\Gamma(\Sigma_b^{(*)\,-}\to\Lambda_b^0\pi^-)$.  To get the values for
$\Gamma(\Sigma_b^{(*)\,0}\to\Lambda_b^0\pi^0)$ and
$\Gamma(\Sigma_b^{(*)\,-}\to\Lambda_b^0\pi^-)$, we use
$g_{\Sigma_b^{(*)\,+}\Lambda_b^0\pi^+}$ and make the appropriate mass
changes in the rest of the factors in
Eqs.~(\ref{eq:gsiglam}-\ref{eq:gsig*lam}).  Our results for the decay
widths are in good agreement with the estimation by the CDF
Collaboration.  We systematically get larger results than in
Ref.~\cite{hwang07} by some $17-38$\% depending on the transition. In
Ref.~\cite{guo08} no attempt is made to get the widths for individual
isospin states. Their theoretical uncertainties result from the
unknown parameters in the model, the scalar and vector diquark masses
and the parameter that describes the confinement interaction between
the heavy quark and the light diquark. Our results lie in the lower
part of the interval determined in Ref.~\cite{guo08}.  Our results are
also some $10-20$\% larger than the ones obtained in
Refs.~\cite{Faessler:2010zzc, Limphirat:2010zz} using the $^3P_0$
production model. In this latter case the results depend on the
parameters that control the strength of the $^3P_0$ creation vertex
and the size of the baryons, which in Refs.~\cite{Faessler:2010zzc,
Limphirat:2010zz} are fitted to reproduce the strange sector
$\Sigma(1385)\to\Lambda\pi$ decay.  Two different fits are quoted in
Ref.~\cite{Limphirat:2010zz}. Much smaller results are obtained in the
$^3P_0$ calculation of Ref.~\cite{Chen:2007xf} where a different size
parameter and a different strength are used.  In all three $^3P_0$
calculations the same size parameter is used to describe the relative
motion of the two light quarks and the relative motion of the heavy
quark with respect to the center of mass of the two light quark
system.  As shown in Figure 4 of Ref.~\cite{Hernandez:2008ej}, those
sizes can be significantly different for $b$-baryons. A mild
dependence on the size parameters is claimed in
Ref.~\cite{Chen:2007xf} though. When compared to the LCSR calculations in 
Refs.~\cite{azizi09} and \cite{azizi11} we find our $g_{\Sigma_b\Lambda_b\pi}$
decay constant is a factor 2.2 and
3.4 larger respectively. The decay constants in Refs.~\cite{azizi09,azizi11} would give rise to rather small
decay widths when compared to other determinations. The  value of
$\Gamma(\Sigma_b\to\Lambda_b\pi)=3.93\pm1.5\,$MeV quoted in Ref.~\cite{azizi09}
is a factor 3.3 larger than one would expect from their value for the decay constant
due to the approximation $(M_{\Sigma_b}-M_{\Lambda_b})^2-m_\pi^2\to
(M_{\Sigma_b}-M_{\Lambda_b})^2$, used in their Eq.~(24), and the approximation
$|\vec q\,|=\frac1{2M_{\Sigma_b}}{\lambda^{1/2}
(M_{\Sigma_b}^2,M_{\Lambda_b}^2,m_\pi^2)}
\to\frac{M^2_{\Sigma_b}-M^2_{\Lambda_b}}{2M_{\Sigma_b}}$ used in their
evaluation of the final momentum. Those approximations are not good because
of the small mass difference between the initial and final
baryons. The
agreement with the LCSR calculation is better for the 
$g_{\Sigma^*_b\Lambda_b\pi}$ coupling constant, being our value larger than the
central value obtained in Ref.~\cite{azizi11-2} by a factor 1.3. This implies
however that our predictions for the decay widths are some 70\% larger.

 \begin{table}[t]
\begin{center}
\begin{tabular}{l|c c c c}
\hline\hline\\
 & \hspace{.25cm}$g_{\Sigma_b^{+}\Lambda_b^0\pi^+}$\hspace{.25cm} &
 \hspace{.25cm}$\Gamma(\Sigma_b^{+}\to\Lambda_b^0\pi^+)$ \hspace{.25cm} &
 \hspace{.25cm} $\Gamma(\Sigma_b^{0}\to\Lambda_b^0\pi^0)$ \hspace{.25cm} &
 \hspace{.25cm}$\Gamma(\Sigma_b^{-}\to\Lambda_b^0\pi^-)$ \hspace{.25cm} 
 \\
 &&[MeV]&[MeV]&[MeV]\\
\hline
This work 
& $51.4$&$6.0$ & $6.6^\P$
&$7.7$\\
\hline
\cite{:2007rw,Korner:1994nh}&&$\approx 7^\dagger$&&$\approx 7^\dagger$\\
\cite{hwang07} &&4.35&$5.65^\ddag$&5.77\\
\cite{guo08}&&$6.73- 13.45$&$6.73- 13.45$&$6.73- 13.45$\\
\cite{Chen:2007xf}&&3.5&&4.7\\
\cite{Faessler:2010zzc}&&&&6.7\\
\cite{Limphirat:2010zz}&&4.82,4.94&&6.31,6.49\\
\cite{azizi09}&$23.5\pm4.9$\\
\cite{azizi11}&$15.0\pm4.9$\\
\hline\hline\\
 & \hspace{.25cm}$g_{\Sigma_b^{*\,+}\Lambda_b^0\pi^+}$\hspace{.25cm} &
 \hspace{.25cm}$\Gamma(\Sigma_b^{*\,+}\to\Lambda_b^0\pi^+)$ \hspace{.25cm} &
 \hspace{.25cm} $\Gamma(\Sigma_b^{*\,0}\to\Lambda_b^0\pi^0)$ \hspace{.25cm} &
 \hspace{.25cm}$\Gamma(\Sigma_b^{*\,-}\to\Lambda_b^0\pi^-)$ \hspace{.25cm} 
 \\
 &&[MeV]&[MeV]&[MeV]\\
\hline
This work 
& $87.6$&$11.0$ & $12.1^\ddag$
&$13.2$\\
\hline

\cite{:2007rw,Korner:1994nh}&&$\approx 13^\dagger$&&$\approx 13^\dagger$\\
\cite{hwang07} &&8.50&$10.20^\ddag$&10.44\\
\cite{guo08}&&$10.00- 17.74$&$10.00- 17.74$&$10.00- 17.74$\\
\cite{Chen:2007xf}&&7.5&&9.2\\
\cite{Faessler:2010zzc}&&&&12.3\\
\cite{Limphirat:2010zz}&&9.68,10.06&&11.81,12.34\\
\cite{azizi11-2}&$67\pm12\,^{\S}$\\

\hline\hline
 \end{tabular}
\end{center}
\caption{ Coupling constants $g_{\Sigma_{b}^{(*)\,+}\Lambda_b^0\pi^+}$
and total widths $\Gamma(\Sigma_b^{(*)\,+}\to\Lambda_b^0\pi^+)$,
$\Gamma(\Sigma_b^{(*)\,0}\to\Lambda_b^b\pi^0)$ and
$\Gamma(\Sigma_b^{(*)\,-}\to\Lambda_b^0\pi^-)$ (See text for
details). The result with a $^\P$ is an estimation using the $\Sigma_b^0$ mass prediction
obtained in the HQET evaluation of Ref~\cite{Hwang:2008dj}. Results with a $\ddag$ are estimations obtained assuming a
mass given by
$M_{\Sigma_b^{(*)\,0}}=(M_{\Sigma_b^{(*)\,+}}+M_{\Sigma_b^{(*)\,-}})/2$. 
Results with a $\dagger$ are estimations by the CDF
Collaboration~\cite{:2007rw} based on the expected masses for the
$\Sigma_b^{(*)}$ baryons and the HQET predictions in
Ref.~\cite{Korner:1994nh}. Result with a $\S$ our estimation from the
value given in Ref.~\cite{azizi11-2} where  the
coupling constant is defined differently.}
\label{tab:siglam}
\end{table}
To end this discussion we compare the present results for the
dimensionless coupling constants $\frac{f_\pi
g_{\Sigma_b^{+}\Lambda_b^0\pi^+}}{M_{\Sigma_b^{+}}+M_{\Lambda_b^0}}$
and $\frac{f_\pi
g_{\Sigma_b^{*\,+}\Lambda_b^0\pi^+}}{2M_{\Lambda_b^0}}$ with the
corresponding ones in the charm sector that we obtained in
Ref.~\cite{Albertus:2005zy}. According to HQS one would expect them to
be equal to the extent that one can consider the $b$ and $c$ quark
masses heavy enough. What we get is
\begin{eqnarray}
\frac{f_\pi\, g_{\Sigma_b^{+}\Lambda_b^0\pi^+}}{M_{\Sigma_b^{+}}+M_{\Lambda_b^0}}=
0.586&\approx&0.598
=\frac{f_\pi\, g_{\Sigma_c^{++}\Lambda_c^+\pi^+}}{M_{\Sigma_c^{++}}+M_{\Lambda_c^+}}\nonumber\\
\frac{f_\pi\,
g_{\Sigma_b^{*\,+}\Lambda_b^0\pi^+}}{2M_{\Lambda_b^0}}=1.02&\approx&1.03=
\frac{f_\pi\, g_{\Sigma_c^{*\,++}\Lambda_c^+\pi^+}}{2M_{\Lambda_c^+}}
 \end{eqnarray}
that coincide at the level of 1-2\%. 

We have obtained accurate predictions of the widths for the
bottom-baryon decays $\Sigma_b\to \Lambda_b\, \pi$ and
$\Sigma_b^{*}\to \Lambda_b\, \pi$ within a constituent quark model
framework. We have used wave functions constrained by HQS and that
were obtained after solving the non-relativistic three-body problem
with the help of a simple variational ansatz. The quality of our wave
functions has been tested in the study of the $\Lambda_b$ and $\Xi_b$
semileptonic decays performed in Ref.~\cite{Albertus:2004wj}, with our
results for partially integrated decay widths being in agreement with
lattice QCD data by the UKQCD Collaboration~\cite{HB-Lattice}. To
evaluate the pion emission amplitude, we have made use of PCAC from
the analysis of weak current matrix elements. A similar procedure was
carried out in \cite{Albertus:2005vd} to evaluate the strong coupling
constants $g_{B^* B \pi}$ and $g_{D^* D \pi}$ that turned out to be in
agreement with the experimental determination\footnote{From the $D^*\to D\pi$
width.} of the latter constant and with lattice
results~\cite{Abada:2003un} for the former one.  Finally, we have
explicitly shown that the corresponding strong couplings $g_{\Sigma_Q
\Lambda_Q\pi}$ and $g_{\Sigma^*_Q \Lambda_Q\pi}$ scale like $m_Q$,
with $Q=b$ or $c$, being the corrections to the infinite mass limit
predictions unexpectedly small, and certainly much smaller than those
found in the the meson sector~\cite{Abada:2003un} .

\begin{acknowledgments}
  This research was supported by DGI and FEDER funds, under contracts
  FIS2006-03438, FIS2008-01143/FIS, FPA2010-21750-C02-02, and the Spanish
  Consolider-Ingenio 2010 Programme CPAN (CSD2007-00042),  by Generalitat
  Valenciana under contract PROMETEO/20090090 and by the EU
  HadronPhysics2 project, grant agreement no. 227431. 
\end{acknowledgments}

\appendix
\section{Description of baryon states and expressions for the 
$A^{1/2,1/2}_{BB',\, \mu}$  weak matrix elements}
The state of a heavy baryon $B$ 
with three-momentum $\vec{P}$ and spin projection $s$ 
 in its center of mass is given as
\begin{eqnarray}
\label{wf}
&&\hspace{-1cm}\left|{B,s\,\vec{P}}\,\right\rangle_{NR}
=\sqrt{2E}\int d^{\,3}Q_1 \int d^{\,3}Q_2\ \frac{1}{\sqrt2}\sum_{\alpha_1,\alpha_2,\alpha_3}
\hat{\psi}^{(B,s)}_{\alpha_1,\alpha_2,\alpha_3}(\,\vec{Q}_1,\vec{Q}_2\,)
\ \frac{1}{(2\pi)^3\ \sqrt{2E_{f_1}2E_{f_2}
2E_{f_3}}}\nonumber\\ 
&&\hspace{3cm}
\times\left|\ \alpha_1\
\vec{p}_1=\frac{m_{f_1}}{\overline{M}}\vec{P}+\vec{Q}_1\ \right\rangle
\left|\ \alpha_2\ \vec{p}_2=\frac{m_{f_2}}{\overline{M}}\vec{P}+\vec{Q}_2\ \right\rangle
\left|\ \alpha_3\ \vec{p}_3=\frac{m_{f_3}}{\overline{M}}\vec{P}-\vec{Q}_1
-\vec{Q}_2\ \right\rangle
 \end{eqnarray}
where the normalization factor $\sqrt{2E}$ has been introduced for
later convenience. $\alpha_j$ represents the quantum numbers of spin
$s$, flavor $f$ and color $c$ ($\alpha_j\equiv(s_j,f_j,c_j)$) of the
{\it j-th} quark, while $(E_{f_j},\,\vec{p}_j)$ and $m_{f_j}$
represent its four--momentum and mass. $\overline{M}$ stands for
$\overline{M}=m_{f_1}+m_{f_2}+m_{f_3}$.  In our case we choose the
third quark to be the $b$ quark while the first two will be the light
ones.  The normalization of the quark states is $ \left\langle\
\alpha^{\prime}\ \vec{p}^{\ \prime}\,|\,\alpha\ \vec{p}\,
\right\rangle=\delta_{\alpha^{\prime},\ \alpha}\, (2\pi)^3\,
2E_f\,\delta( \vec{p}^{\ \prime}-\vec{p}\,) $.  Besides,
$\hat{\psi}^{\,(B,s)}_{\alpha_1,\alpha_2,\alpha_3}
(\,\vec{Q}_1,\vec{Q}_2\,)$ is the non-relativistic momentum space wave
function for the internal motion, being $\vec{Q}_1$ and $\vec{Q}_2$
the momenta conjugate to the relative positions $\vec{r}_1$ and
$\vec{r}_2$ of the two light quarks with respect to the heavy
one. This wave function is antisymmetric under the simultaneous
exchange $\alpha_1\longleftrightarrow \alpha_2, \vec{Q}_1
\longleftrightarrow \vec{Q}_2 $, being also antisymmetric under an
overall exchange of the color degrees of freedom. It is normalized
such that
\begin{equation}
\int d^{\,3}Q_1 \int d^{\,3}Q_2\ \sum_{\alpha_1,\alpha_2,\alpha_3}
\left(\hat{\psi}^{(B,s')}_{\alpha_1,\alpha_2,\alpha_3}(\,\vec{Q}_1,\vec{Q}_2\,)\right)^*
\hat{\psi}^{(B,s)}_{\alpha_1,\alpha_2,\alpha_3}(\,\vec{Q}_1,\vec{Q}_2\,)
=\delta_{s',\, s}
\end{equation}
and, thus,  the normalization of our non-relativistic baryon states is 
\begin{equation}
{}_{\stackrel{}{NR}}\left\langle\, {B,s'\,\vec{P}^{\,\prime}}\,|\,{B,s
\,\vec{P}}\,\right\rangle_{NR}
=\delta_{s',\,s}\,(2\pi)^3\,2E\,\delta(\vec{P}^{\,\prime}-\vec{P}\,)
\end{equation}
For the particular case of ground state $\Lambda_b$, $\Sigma_b$, and
$\Sigma^{*}_b$, we assume the orbital angular momentum to be zero. We
will also take advantage of HQS and assume the light--degrees of
freedom quantum numbers are well defined (For quantum numbers see, for
instance, Table 1 in Ref~\cite{albertus04}). In that case we
have\footnote{We only give the wave function for the baryons involved
in $\pi^+$ emission. Wave functions for other isospin states of the
same baryons are easily constructed.}
\begin{eqnarray}
\hat{\psi}^{\,(\Lambda^0_b, s)}_{\alpha_1,\,\alpha_2,\,\alpha_3}(\,\vec{Q}_1,\vec{Q}_2\,)
&=&\frac{\varepsilon_{c_1\,c_2\,c_3}}{\sqrt{3}!}
\ (1/2,1/2,0;s_1,s_2,0)\nonumber\\
&&\hspace{2cm}\times
\frac{\delta_{f_3,\,b}\,\delta_{s_3,\,s}}{\sqrt2}
\left(
\delta_{f_1,\,u}\,\delta_{f_2,\,d}\,\widetilde{\phi}^{\Lambda_b^0}_{u,\,d,\,b}(\,\vec{Q}_1,\vec{Q}_2\,)\ 
-\delta_{f_1,\,d}\,\delta_{f_2,\,u}\,\widetilde{\phi}^{\Lambda_b^0}_{d,\,u,\,b}(\,\vec{Q}_1,\vec{Q}_2\,)\
\right)  
 \nonumber\\
\hat{\psi}^{\,(\Sigma^{+}_b,s)}_{\alpha_1,\,\alpha_2,\,\alpha_3}(\,\vec{Q}_1,\vec{Q}_2\,)
&=&\frac{\varepsilon_{c_1\,c_2\,c_3}}{\sqrt{3}!}\
\ \widetilde{\phi}^{\Sigma_b^+}_{u,\,u,\,b}(\,\vec{Q}_1,\vec{Q}_2\,)\ 
\delta_{f_1,\,u}\, \delta_{f_2,\,u}\, \delta_{f_3,\,b}
\sum_m (1/2,1/2,1;s_1,s_2,m)\ 
(1,1/2,1/2;m,s_3,s)\nonumber\\
\hat{\psi}^{\,(\Sigma^{*\,+}_b,s)}_{\alpha_1,\,\alpha_2,\,\alpha_3}(\,\vec{Q}_1,\vec{Q}_2\,)
&=&\frac{\varepsilon_{c_1\,c_2\,c_3}}{\sqrt{3}!}\
\ \widetilde{\phi}^{\Sigma_b^*{+}}_{u,\,u,\,b}(\,\vec{Q}_1,\vec{Q}_2\,)\ 
\delta_{f_1,\,u}\,\delta_{f_2,\,u}\, \delta_{f_3,\,b}
\ 
\sum_m (1/2,1/2,1;s_1,s_2,m)\ 
(1,1/2,3/2;m,s_3,s)\nonumber\\
\end{eqnarray}
Here $\varepsilon_{c_1 c_2 c_3}$ is the fully antisymmetric tensor on
color indices being $\frac{\varepsilon_{c_1 c_2 c_3}}{\sqrt{3!}}$ the
antisymmetric color wave function, the $(j_1,j_2,j;m_1,m_2,m_3)$ are
Clebsch-Gordan coefficients and the
$\widetilde{\phi}^{B}_{f_1,\,f_2,\,f_3}(\,\vec{Q}_1, \vec{Q}_2\,)$ are
the Fourier transform of the corresponding normalized coordinate space
wave functions obtained in Ref.~\cite{albertus04} using the AL1
potential or Ref.~\cite{semay94,silvestre96}.  Their dependence on
momenta is through $|\vec{Q}_1|$, $|\vec{Q}_2|$ and
$\vec{Q}_1\cdot\vec{Q}_2$ alone, and they are symmetric under the
simultaneous exchange $f_1\longleftrightarrow f_2, \vec{Q}_1
\longleftrightarrow \vec{Q}_2 $. \\
Within the model we evaluate the $A^{1/2,1/2}_{BB',\, \mu}$ as
\begin{eqnarray}
A_{\Sigma_b^{(*)\,+}\Lambda_b^0
,\ \mu}^{1/2,1/2}={}_{\stackrel{}{NR}}\left\langle \Lambda_b^0,\
1/2\ -|\vec{q}|\,\vec{k}\ \left|\  J_{A\,\mu}^{d\, u}(0)\ \right|
\,\Sigma_b^{(*)\,+},\ 1/2\ \vec{0}\,\right\rangle_{NR\ non-pole}
\end{eqnarray}
Using one-body current operators their expressions are given by
\begin{eqnarray}
A^{1/2,1/2}_{\Sigma_b^{+}\Lambda_b^0,\, \mu}
&=&\frac{\sqrt2}{\sqrt3}
\sqrt{2M_{\Sigma_b^+}2E_{\Lambda_b^0}}\int d^{\,3}Q_1\ d^{\,3}Q_2\ 
\phi^{\Sigma_b^+}_{u,u,b}(\vec{Q}_1,\vec{Q}_2)
\left(
\phi^{\Lambda_b^0}_{d,u,b}(\vec{Q}_1-\frac{m_u+m_b}{\overline{M}_{\Lambda_b^0}}\, |
\vec{q}\,|\,\vec{k},\ \vec{Q}_2+\frac{m_u}{\overline{M}_{\Lambda_b^0}}\, |
\vec{q}\,|\,\vec{k} )
\right)^*\nonumber\\
&&\times\ \sum_{s_1}\,(1/2,1/2,1;s_1,-s_1,0)\,(1/2,1/2,0;s_1,-s_1,0)
\ \frac{
\overline{u}_{d\ s_1}(\vec{Q}_1-|
\vec{q}\,|\,\vec{k} )\ \gamma_\mu\gamma_5\ u_{u\, s_1}(\vec{Q}_1)}{\sqrt{2E_d(|\vec{Q}_1-|
\vec{q}\,|\,\vec{k} |)2E_u(|\vec{Q}_1|)}}\ 
\end{eqnarray}
%
%
%
where the quark Dirac spinors are normalized to twice the energy. For
$\mu=0,3$ we get the final expressions

\begin{eqnarray}
&&\hspace{-2cm}A^{1/2,1/2}_{\Sigma_b^{+}\Lambda_b^0,\, 0}
=\frac{\sqrt2}{\sqrt3}
\sqrt{2M_{\Sigma_b^+}2E_{\Lambda_b^0}}\int d^{\,3}Q_1\ d^{\,3}Q_2\ 
\phi^{\Sigma_b^+}_{u,u,b}(\vec{Q}_1,\vec{Q}_2)
\left(
\phi^{\Lambda_b^0}_{d,u,b}(\vec{Q}_1-\frac{m_u+m_b}{\overline{M}_{\Lambda_b^0}}\, |
\vec{q}\,|\,\vec{k},\ \vec{Q}_2+\frac{m_u}{\overline{M}_{\Lambda_b^0}}\, |
\vec{q}\,|\,\vec{k} )
\right)^*\nonumber\\
&&\times\ \sqrt{\frac{\left(E_d(|\vec{Q}_1-|
\vec{q}\,|\,\vec{k} |)+m_d\right)\left(E_u(|\vec{Q}_1|)+m_u\right)}{{2E_d(|\vec{Q}_1-|
\vec{q}\,|\,\vec{k} |)2E_u(|\vec{Q}_1|)}}}\ 
\left(
\frac{Q_1^z}{E_u(|\vec{Q}_1|)+m_u}+
\frac{Q_1^z-|\vec{q}\,|}{E_d(|\vec{Q}_1-|
\vec{q}\,|\,\vec{k} |)+m_d}
\right)
\end{eqnarray}
\begin{eqnarray}
&&\hspace{-1.5cm}A^{1/2,1/2}_{\Sigma_b^{+}\Lambda_b^0,\, 3}
=-\frac{\sqrt2}{\sqrt3}
\sqrt{2M_{\Sigma_b^+}2E_{\Lambda_b^0}}\int d^{\,3}Q_1\ d^{\,3}Q_2\ 
\phi^{\Sigma_b^+}_{u,u,b}(\vec{Q}_1,\vec{Q}_2)
\left(
\phi^{\Lambda_b^0}_{d,u,b}(\vec{Q}_1-\frac{m_u+m_b}{\overline{M}_{\Lambda_c^+}}\, |
\vec{q}\,|\,\vec{k},\ \vec{Q}_2+\frac{m_u}{\overline{M}_{\Lambda_c^+}}\, |
\vec{q}\,|\,\vec{k} )
\right)^*\nonumber\\
&&\hspace{.3cm}\times\ \sqrt{\frac{\left(E_d(|\vec{Q}_1-|
\vec{q}\,|\,\vec{k} |)+m_d\right)\left(E_u(|\vec{Q}_1|)+m_u\right)}{{2E_d(|\vec{Q}_1-|
\vec{q}\,|\,\vec{k} |)2E_u(|\vec{Q}_1|)}}}
\left(1+\frac{ 2(Q_1^z)^2-|\vec{Q}_1|^2-Q_1^z|
\vec{q}\,|}{\left(E_d(|\vec{Q}_1-|
\vec{q}\,|\,\vec{k} |)+m_d\right)\left(E_u(|\vec{Q}_1|)+m_u\right)}
\right)
\end{eqnarray}
\vspace{1cm}\\ 
For $A^{1/2,1/2}_{\Sigma_b^{*\,+}\Lambda_b^0,\, \mu}$
we get similar relations with an extra $-\sqrt2$ factor.
%
%
%
%
%
%
%
%
%
%
%

\end{document}